\documentclass{emulateapj}
\usepackage{graphicx}
\usepackage{subfigure}
\usepackage{mathrsfs,amssymb}
\usepackage{amsmath}
\usepackage{natbib}
\usepackage{color}
\usepackage{ulem}
\usepackage{txfonts}

\def\gtsima{$\; \buildrel > \over \sim \;$}
\def\ltsima{$\; \buildrel < \over \sim \;$}
\def\gsim{\lower.5ex\hbox{\gtsima}}
\def\lsim{\lower.5ex\hbox{\ltsima}}
\def\simgt{\lower.5ex\hbox{\gtsima}}
\def\simlt{\lower.5ex\hbox{\ltsima}}
\def\simpr{\lower.5ex\hbox{\prosima}}
\def\mean#1{\left< #1 \right>}

 \newcommand*\oline[1]{%
  \vbox{%
    \hrule height 0.5pt
    \kern0.25ex
    \hbox{%
      \kern-0.1em
      \ifmmode#1\else\ensuremath{#1}\fi
      \kern-0.1em
    }
  }
}

\shorttitle{Frontier Fields constraints on photoionization feedback}
\shortauthors{M. Castellano et al.}
\begin{document} 

   \title{Constraints on photoionization feedback from number counts of ultra-faint high-redshift galaxies in the Frontier Fields}

\author{M. Castellano\altaffilmark{1}, B. Yue\altaffilmark{2}, A. Ferrara\altaffilmark{2,3}, E. Merlin\altaffilmark{1}, A. Fontana\altaffilmark{1}, R. ~Amor\'{\i}n \altaffilmark{1}, A. Grazian\altaffilmark{1}, E. ~M\'armol-Queralto\altaffilmark{4}, M.~J.~Micha{\l}owski\altaffilmark{4}, A. ~Mortlock\altaffilmark{4}, D. Paris\altaffilmark{1}, S. ~Parsa\altaffilmark{4}, S. Pilo\altaffilmark{1},  P. ~Santini\altaffilmark{1}}

\altaffiltext{1}{INAF - Osservatorio Astronomico di Roma, Via Frascati 33, I - 00078 Monte Porzio Catone (RM), Italy}
\altaffiltext{2}{Scuola Normale Superiore, Piazza dei Cavalieri 7, I-56126 Pisa, Italy}
\altaffiltext{3}{Kavli IPMU (WPI), Todai Institutes for Advanced Study, the University of Tokyo, Japan}
\altaffiltext{4}{SUPA, Scottish Universities Physics Alliance, Institute for Astronomy, University of Edinburgh, Royal Observatory, Edinburgh, EH9 3HJ, U.K.}
\email{marco.castellano\char64oa-roma.inaf.it}

 \begin{abstract}
We exploit a sample of ultra-faint high-redshift galaxies (demagnified HST $H_{160}$ magnitude $>30$) in the Frontier Fields clusters A2744 and M0416 to constrain  a theoretical model for the UV luminosity function (LF) in the presence of photoionization feedback. The objects have been selected on the basis of accurate photometric redshifts computed from multi-band photometry including 7 HST bands and deep $K_s$ and IRAC observations. Magnification is computed on an object-by-object basis from all available lensing models of the two clusters. We take into account source detection completeness as a function of luminosity and size, magnification effects and systematics in the lens modeling of the clusters under investigation. We find that our sample of high-$z$ galaxies constrain the cut-off halo circular velocity below which star-formation is suppressed by photo-ionization feedback to $v_c^{\rm cut} < 50$ km s$^{-1}$. This circular velocity corresponds to a halo mass of $\approx5.6\times10^9~M_\odot$ and  $\approx2.3\times10^9~M_\odot$ at $z=5$ and 10 respectively: higher mass halos can thus sustain continuous star formation activity without being quenched by external ionizing flux. More stringent constraints are prevented by the uncertainty in the modeling of the cluster lens, as embodied by systematic differences among the lens models available.
\end{abstract}

\keywords{dark ages, reionization, first stars --- galaxies: high-redshift}

   \maketitle

\section{Introduction}

The investigation of the reionization process and of the earliest phases of galaxy evolution are deeply connected. Star-forming galaxies are currently believed to be the sources of reionizing photons, with the bulk of the ionizing flux generated by objects at the faint end of the luminosity function (LF) \citep[e.g.][]{Bouwens2015b,Robertson2015,Finkelstein2015,Castellano2016}, although we cannot yet rule out a  contribution from bright star-forming galaxies \citep{Sharma2016} or AGNs \citep[e.g.][]{Giallongo2015,Yoshiura2016}. Our understanding of the reionization epoch is currently limited by a poor knowledge on key physical quantities such as the escape fraction of ionizing photons \citep[e.g.][]{Khaire2015}, the intrinsic ionizing budget \citep[e.g.][]{Stanway2016,Ma2016} and the cut-off of the UV LF at faint magnitudes \citep[e.g.][]{Bouwens2015b}. The investigation of the LF cut-off is of particular interest because of its relation with star-formation and feedback processes in low mass halos. 

Faint galaxies are hosted by low mass halos with shallow gravitational potentials: in the presence of an external ionizing flux their gas could be evaporated and the star formation quenched (e.g. \citealt{2008MNRAS.390.1071M,2013MNRAS.432L..51S,2013MNRAS.432.3340S,2013MNRAS.428..154H}). This may eventually result in a reduction of the number of ionizing photons they emit, thus questioning their role in ionizing the IGM. The reionization process and its sources interplay with each other.  Until now there is a lack of direct observations of such a picture.  

The gravitational lensing provides us with the opportunity to investigate such faint galaxy populations. The Frontier Fields (FF) Survey provides the ideal context for such an investigation. The FF survey is an HST observing program targeting six galaxy clusters, and six parallel pointings at depths comparable to the Hubble Ultra Deep Field one. Thanks to magnification effects the FF survey enables the study of galaxies as intrinsically faint as those that will be detected by JWST in blank fields. In principle, the effect of feedback can be investigated through a direct derivation of the UV LF to look for a  cut-off of the galaxy number density distribution. Here we take an alternative and more powerful approach described in \citet{2014MNRAS.443L..20Y} (Y14 hereafter), namely a comparison between the observed number counts and those predicted by a theoretical model of formation and evolution of galaxies during the reionization epoch \citep{Yue2016}. As shown by Y14, such an approach is extremely sensible and enables constraints even from limited galaxy samples. We will exploit the technique outlined in Y14 using data from the first two Frontier Fields Abell-2744 (A2744 hereafter) and MACSJ0416.1-2403 (M0416).

Throughout the paper, observed and rest--frame magnitudes are in
the AB system, and we adopt the $\Lambda$-CDM concordance model ($\Omega_m$=0.308, $\Omega_\Lambda$= 0.692, $\Omega_b$=0.048, h=0.678, $\sigma_8$=0.815, $n_s$=0.97, see \citealt{2015arXiv150201589P}).

\section{The Frontier Fields dataset}\label{sect_FFDATA}

We exploit the ASTRODEEP photometric redshift catalogs of A2744 and M0416 from \citet{Castellano2016b} (C16b hereafter) based on the multi-band photometry  presented in \citet{Merlin2016} (M16 hereafter)\footnote{http://www.astrodeep.eu/frontier-fields/}. We summarise here the information most relevant for the present work.

The catalogs include information for 10 passbands: the seven HST bands observed under the FF program (F435W, F606W, F814W, F105W, F125W, F140W and F160W) together with Hawk-I@VLT \textit{$K_s$} band and IRAC 3.6 and 4.5 $\mu$m data. The typical $5\sigma$ depth in 2 PSF-FWHM apertures are $\sim28.5-29.0$ (HST filters), $\sim26.2$ ($K_s$), $\sim25$ (IRAC). The detection is performed on the F160W band ($H_{160}$ hereafter) after applying a procedure (see M16) to remove foreground light both from bright cluster galaxies and the diffuse intra-cluster light (ICL). Low resolution $K_s$ and IRAC images have been processed with \verb|T-PHOT| \citep{Merlin2015}. As shown in M16, this procedure enhances the detection of faint lensed galaxies, especially in the central regions of the clusters.

Photometric redshifts have been measured with six different techniques based on different codes and assumptions (see C16b). The FF sources are then assigned the median of the six available estimates in order to minimize systematics and improve the accuracy. In the cluster fields the typical accuracy found from a comparison with spectroscopic samples is $\sigma_{\Delta z/(1+z)}\sim0.04$. We successfully recover as high-$z$ sources most of the $z>6$ candidates known in the two fields  \citep{Laporte2014,Zitrin2014,Oesch2014,Zheng2014,Atek2015,Coe2015,McLeod2015,Ishigaki2015}.
\begin{figure*}
\centering{
\includegraphics[width=12.5cm]{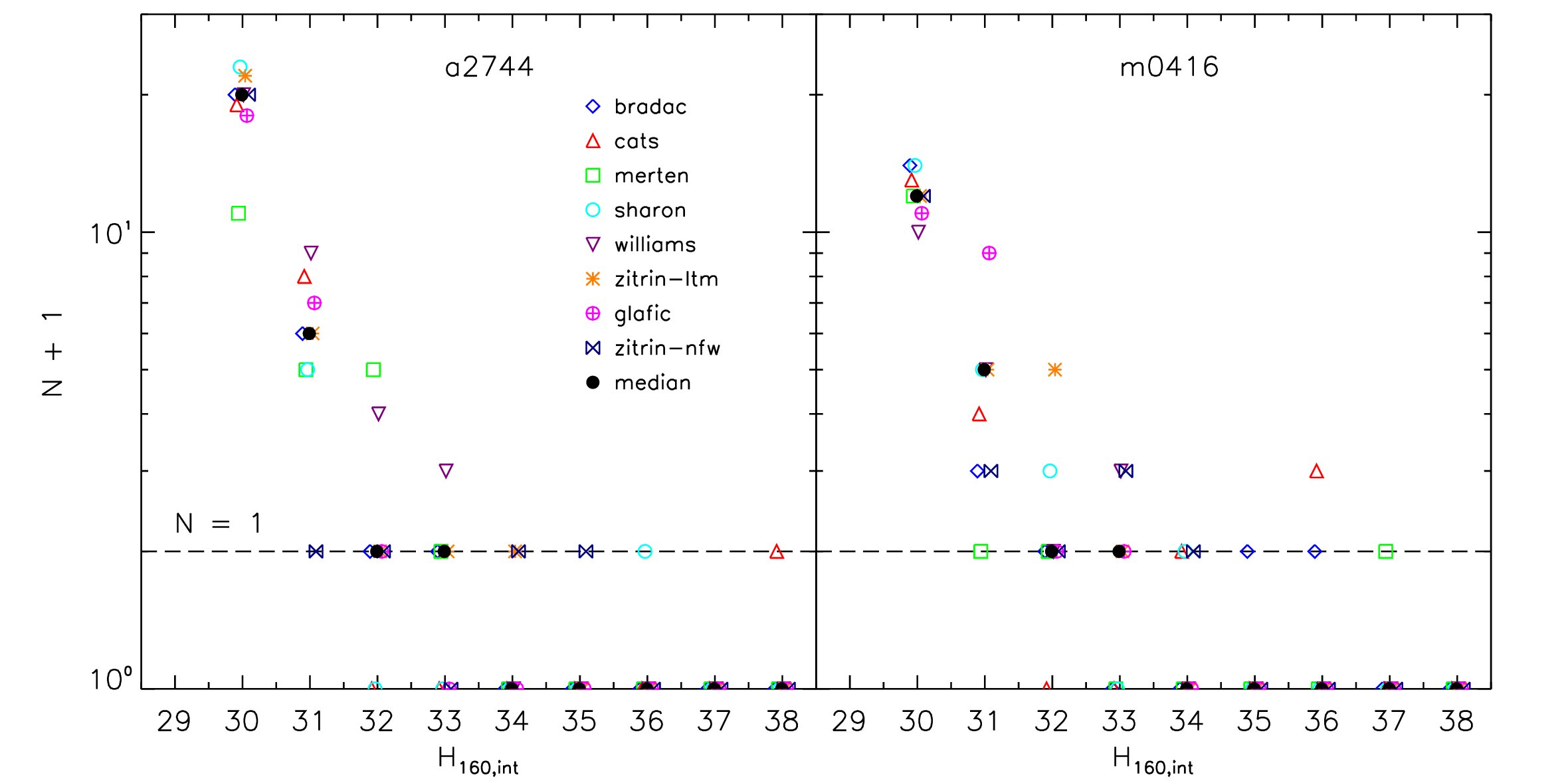}
\includegraphics[width=11.5cm]{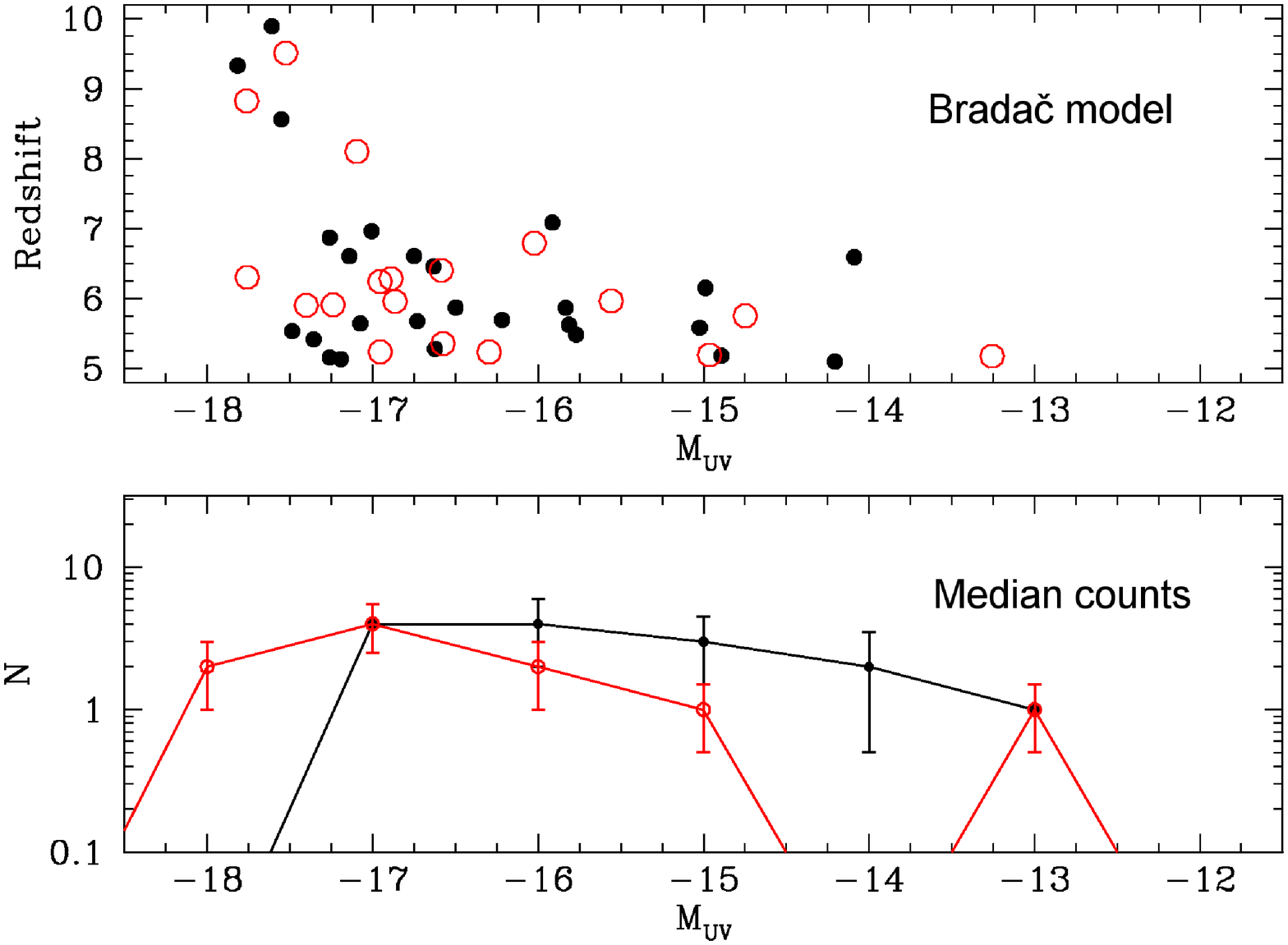}
\caption{\textbf{Top:} the number of galaxies in intrinsic $H_{160,int}$ bins of width 1.0 mag for the clusters A2744 (left) and M0416 (right) respectively. Different symbols refer to different lensing models.  For displaying purpose we slightly shift the x-axes of some models within the magnitude bin. To guide the eye we plot the $N=1$ as horizontal dashed line. \textbf{Bottom:} de-magnified UV rest-frame magnitudes in A2744 (black lines and filled circles) and M0416 (red lines and empty circles): as a function of redshift (top panel) and  distribution (bottom). The Brada\u{c} model is used for reference in the former plot; median among number counts from the eight models is used in the latter (error bars show the interquartile range).}
\label{lensing_models}
}
\end{figure*}
We assign magnification values to sources in our catalogs as estimated from the different lensing models of the two FF fields\footnote{http://www.stsci.edu/hst/campaigns/frontier-fields/Lensing-Models}. We measure for each object the shear and mass surface density values at its position from the relevant maps and we use them to compute the magnification at the source photometric redshift.

\subsection{High-redshift sample}

Following Y14 in this work we will consider objects at $5.0<z<10.0$ and with intrinsic magnitude fainter than $H_{160,int}=30.0$. Such faint sources are typically hosted in halos of mass $\sim10^{8.5}-10^{9.5} M_{\odot}$, corresponding to virial temperatures $T_{vir}\gtrsim1.5\times 10^4$~K, that are likely strongly affected by photoionization feedback \citep[e.g.][Y14]{Dijkstra2004,2008MNRAS.390.1071M}. In Fig.~\ref{lensing_models} we show the number of galaxies with \textit{intrinsic} $H_{160,int}>30$ in magnitude bins with width $\Delta H_{160,int} = 1$ as obtained by demagnifing \textit{observed} magnitudes following the different available lensing models in the two fields. The number of $H_{160,int}>30$ high-$z$ sources ranges from 19 to 32 in the A2744 field, and from 14 to 20 in the M0416 field, depending on the adopted lensing map. We find that most of these are faint $H_{160}\sim$28-29 sources magnified by a factor $\sim5-10$ with only a small fraction of the objects ($\sim10$\%, depending on the model) being selected thanks to an extremely high magnification ($\gtrsim50$). Objects at $z\sim5-7$ constitute the bulk of the sample outnumbering higher redshift sources by a factor of 7-8. The samples selected according to different models typically include different sources, with only about half of the objects being selected by 2 or more models in each field. Despite these differences, the number counts obtained from different lensing models show a similar behavior. In Sect.~\ref{results} we will describe the procedure we exploit to derive theoretical constraints while taking into account these discrepancies between different lensing models.

\subsection{Completeness simulations}\label{completeness}

A critical ingredient for comparing predicted and observed number counts of ultra-faint lensed sources is the detection completeness as a function of the observed $H_{160}$ magnitude. As described in M16 we estimate completeness through imaging simulations with synthetic sources of different magnitudes and sizes. We consider both point-like and exponential profile sources with half-light radius $0.05<R_{h}<1.0$ arcsec, and total magnitude $26.5<H_{160}<30.0$. We simulate $2\times10^5$ sources per field. 
Two hundred mock galaxies each time are placed at random positions in our detection image which is then analysed using the same \textsc{SExtractor} parameters adopted in the real case. After the whole input galaxy population has been analysed we store the tabulated values of the completeness at different magnitudes and $R_{h}$ that will be used in Sect.~\ref{results} for comparing observation to our model. As a reference, the 90\% detection completeness ranges from $H_{160}\sim26.6-26.7$ ($R_{h}$=0.3 arcsec) to $H_{160}\sim27.7-27.8$ in the case of point-sources. 

\section{The model}\label{sect_MODEL}

We compare our observations to a theoretical model for the LF of high-$z$ galaxies in the presence of reionization feedback. The model extends the analytical algorithm described in  \citet{2010ApJ...714L.202T,2013ApJ...768L..37T,2015arXiv150801204M} by additionally including the quenching of star formation activity in low mass halos that are located in ionized regions. Full details are presented in the above mentioned papers and in \citet{Yue2016}, here we briefly summarize the main properties. The model relies on the assumption of a star formation efficiency which is a redshift-independent function of halo mass. Halos with the same final mass can have different luminosities as a result of different mass assembly histories. The star formation efficiency parameter is calibrated from the observed $z=5$ UV LFs and then used to model the LF at higher redshifts on the basis of the halo mass function and the above constructed luminosity - halo mass relations. As pointed out by \citet{2015arXiv150801204M} this approach allows us to reproduce the observed high-$z$ LFs.

The effect of photoionization feedback on the star formation activity depends on the halo mass: 1) SFR is suppressed in halos with circular velocity below a given cut-off value ($v_c^{\rm cut}$) that form in already ionized regions; 2) star-formation can begin in halos with $v_c<v_c^{\rm cut}$ that formed in neutral regions but it is then quenched if their environment is ionized by neighboring galaxies; 3) star-formation proceeds uninterrupted as the host halos are massive enough ($>v_c^{\rm cut}$) all the time.  Using the ``bubble model" based algorithm presented in \citet{2004MNRAS.354..695F} we model the  above three cases to find the probability for a given halo to be located in an ionized bubble large enough to contain at least another persistent galaxy (i.e. always having $>v_c^{\rm cut}$) and revise the halo star formation history described above accordingly. We eventually obtain the UV luminosity and emission rate of ionizing photons of a halo when its mass, formation time, star formation quench time are given. The model has two free parameters, the escape fraction of ionizing photons ($f_{\rm esc}$) and the above mentioned cut-off circular velocity $v_c^{\rm cut}$, that provide a flexible way to model the interconnection between UV background and feedback effects on the star-formation. These two parameters are treated as independent of each other in our model: for each given pair of values we compute the reionization history and the resulting UV LFs in a self-consistent way \citep[see ][for details]{Yue2016}.  

As an effect of reionization feedback, the abundance of galaxies in halos with $v_c<v_c^{\rm cut}$ drops rapidly (although not necessarily monotonically). Interestingly, due to the extremely steep intrinsic UV LF faint-end, even a strong reionization feedback (i.e. high $f_{\rm esc}$ and $v_c^{\rm cut}$) is not enough to make the abundance drop to zero, such that faint galaxies with $v_c<v_c^{\rm cut}$ may still exist and be numerous even after reionization is completed. These galaxies can start their initial star formation activity at the formation time and are then quenched later on. They act as a fossil record of the reionization process: their abundance allows us to constrain the reionization history. Finally, we remark that our model assumes $\Lambda CDM$ cosmology, and the abundance of low mass galaxies is consistently interpreted as affected by feedback effects on star-formation in low mass halos. However, modifications of the initial power spectrum as in WDM cosmologies can also affect number counts at the faintest end in a similar way \citep[e.g.][]{Barkana2001b,Dayal2015,Menci2016}.

\begin{figure*}[!ht]
\centering{
\includegraphics[width=7.5cm]{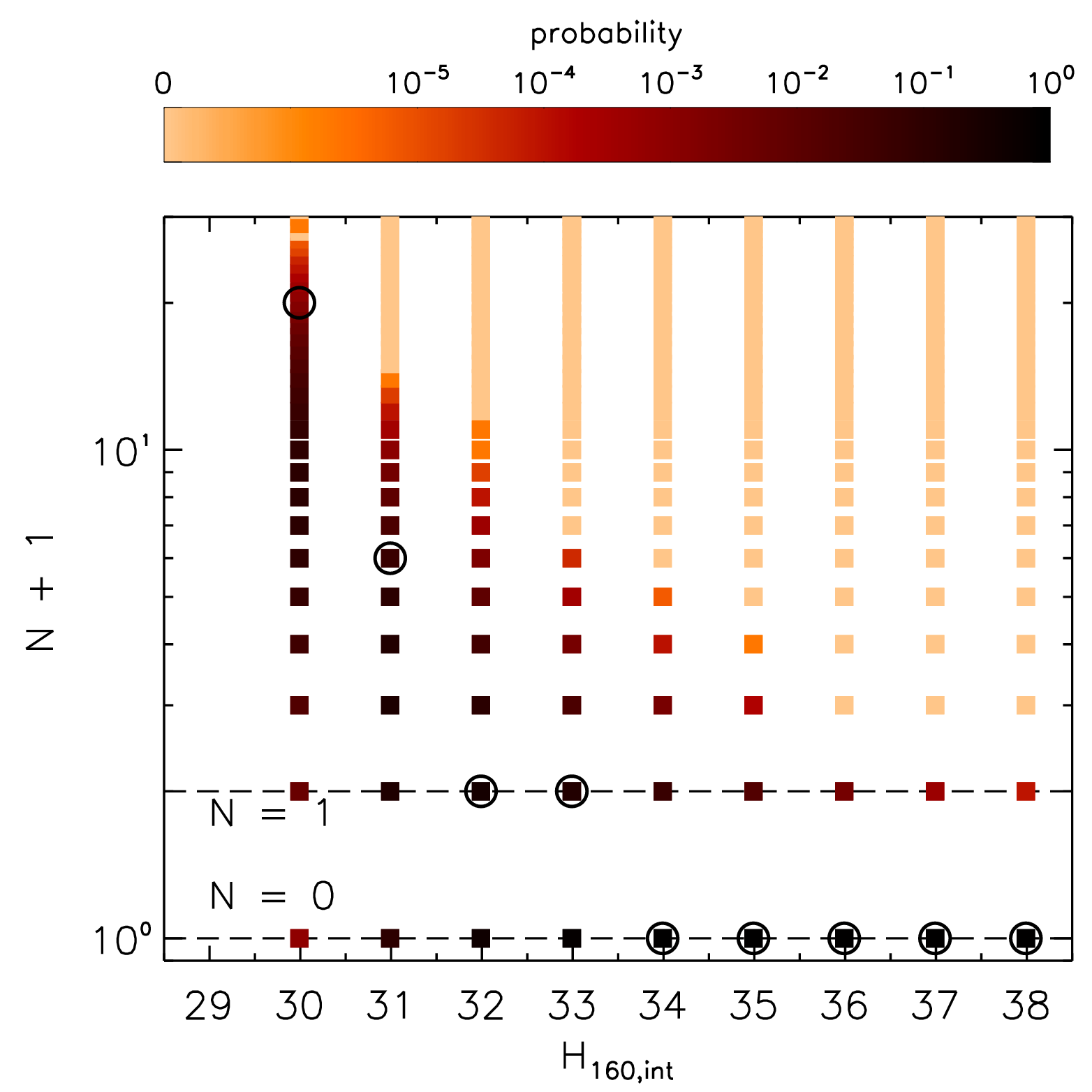}
\includegraphics[width=7.5cm]{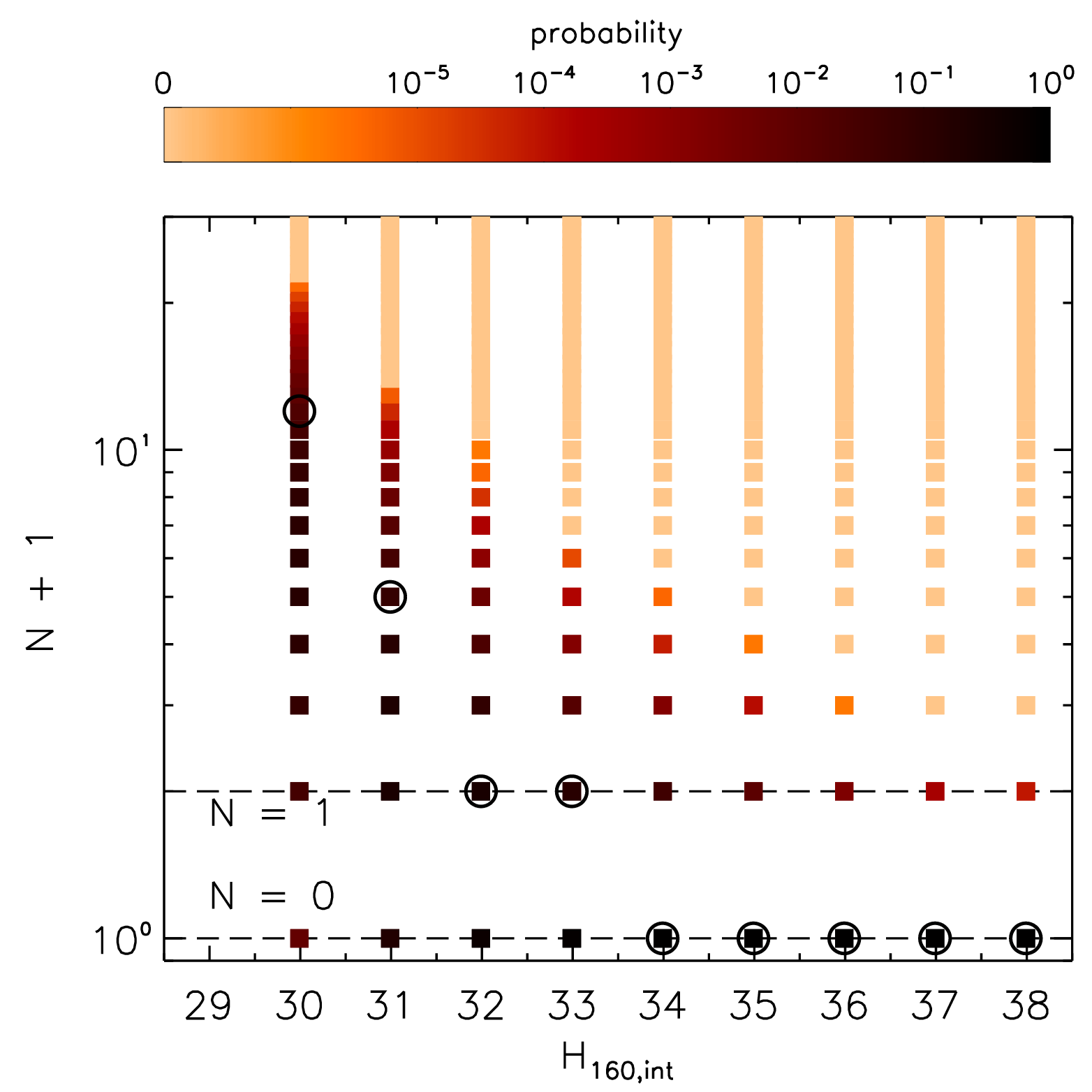}
\caption{ 
The probabilities of observeing various number of galaxies in different magnitude bins for model parameters $f_{\rm esc} = 0.2$ and $v_c^{\rm cut} = 30$~km s$^{-1}$. {\it Left} panel is for the A2744, {\it right} panel is for M0416. The open circles are the median of the observed number counts obtained from different lensing maps.  To guide the eye we plot the $N=1$ and $N=0$ as horizontal dashed lines.
}
\label{probability}
}
\end{figure*}

\section{Constraints on the LF cut-off}\label{results}

In this section we investigate the constraints we can put thanks to the FFs high-$z$ sample on the two free parameters in our model: $f_{\rm esc}$ and $v_c^{\rm cut}$. We exploit Monte Carlo simulations to compute the probabilities to observe different number of galaxies once the LF is given. First, we have the mean number of galaxies in the effective volume behind one pixel in the lensing model,
\begin{equation}
\mean{N}=\Delta\Omega\int dz_s r^2(z_s)\frac{dr}{dz_s}\frac{1}{\mu(z_s)}\int\Phi(L,z_s)dL,
\end{equation}
where $\Delta\Omega$ is the solid angle of this pixel, $r$ is the comoving distance and $\mu$ is the magnification.  We then randomly generate an integer $N$ from a Poisson distribution with mean value $\mean{N}$, so there are $N$ galaxies in hand; each galaxy is assigned a redshift from the probability distribution $\propto d\mean{N}/dz_s$, and then a luminosity from the probability distribution $\propto \Phi(L,z_s)$ where $\Phi(L,z_s)$ is the theoretical LF for a given $f_{\rm esc}$ and $v_c^{\rm cut}$ \citep{Yue2016}. We assign to this galaxy a physical size $R$ that follows a log-normal distribution. As in \citet{2013ApJ...765...68H}, the peak of the distribution is luminosity-dependent, 
\begin{equation}
\bar{R}(L)=R_0\left(\frac{L}{L_0}\right)^\beta.
\end{equation}
We take the best fitting parameters at $z_0=5$ (however parameters at $z_0 = 4$ yield similar results) in their Table 3; namely $R_0 = 1.19$~kpc, $L_0$ corresponding to an absolute magnitude $M_0 = -21$, $\beta = 0.25$. The variance of the distribution is $\sigma_{{\rm log}R} = 0.9/{\rm ln10}$. The physical size at other redshifts is derived from a $\propto(1+z)^{-1}$ evolution. We verified that a different assumption on the evolution of size with redshift, as the $\propto(1+z)^{-0.47}$ found by \citet{Curtis-Lake2016}, does not significantly affect the results. The last step is to include observational incompleteness on the basis of simulations described in Sect.~\ref{completeness}. To this aim we find the apparent magnitudes for the above $N$ galaxies as $H_{160,int} - 2.5{\rm log}(\mu)$ and the observed angular sizes $\sqrt{\mu}\times\theta$, where $H_{160,int}$ is the intrinsic magnitude and $\theta$ is the intrinsic angular size. The predicted number counts are then scaled on the basis of the estimated completeness level for galaxies of the given observed magnitude and size.

After the loop for all lensing pixels that are in the WFC3/HST field of view, {\it one} random realization is completed. For each lensing model we eventually make 30000 realizations.  In the above algorithm, each pixel is treated as independent of each other, so if the effective volume of different pixels overlaps (as in the case of multiple images), galaxies in the overlapped volume are counted more than once. Therefore, when comparing simulated samples with observations, the number of images instead of the number of objects, should be compared, such that we are not interested in determining whether our high-$z$ samples include multiple images of the same observed source.

\begin{figure}[!ht]
\centering{
\includegraphics[width=8.0cm]{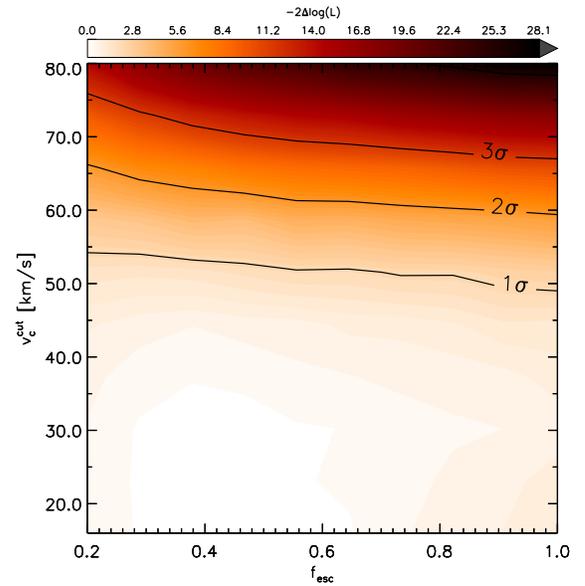}
\caption{Likelihood map for $f_{\rm esc}$ and $v_c^{\rm cut}$ as constrained by the comparison between
the combination of A2744 and M0416 high-redshift samples from all available lensing models and our theoretical model.}
\label{contour_z}
}
\end{figure}
\begin{figure}[!ht]
\centering{
\includegraphics[width=8cm]{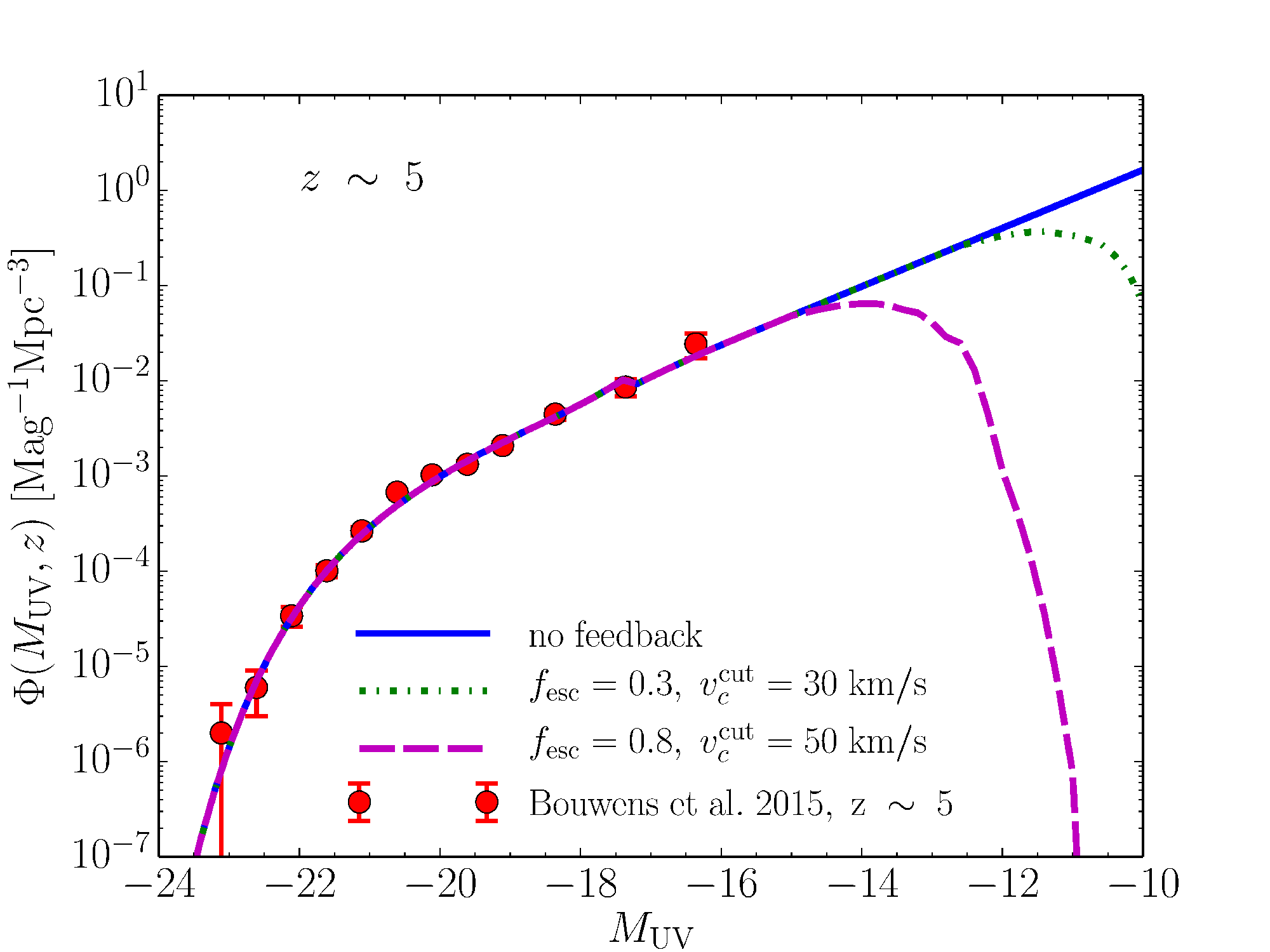}
\includegraphics[width=8cm]{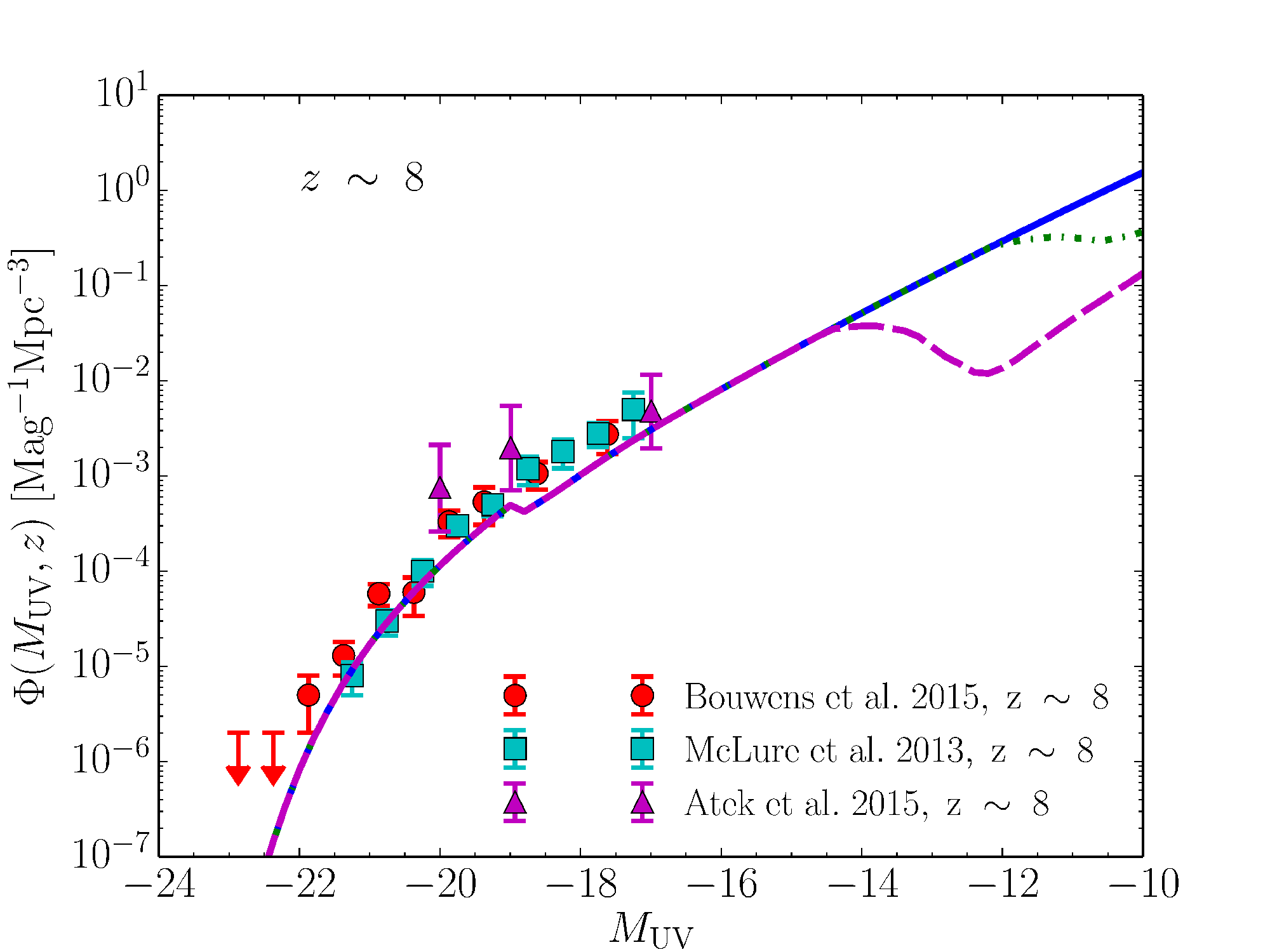}
\includegraphics[width=8cm]{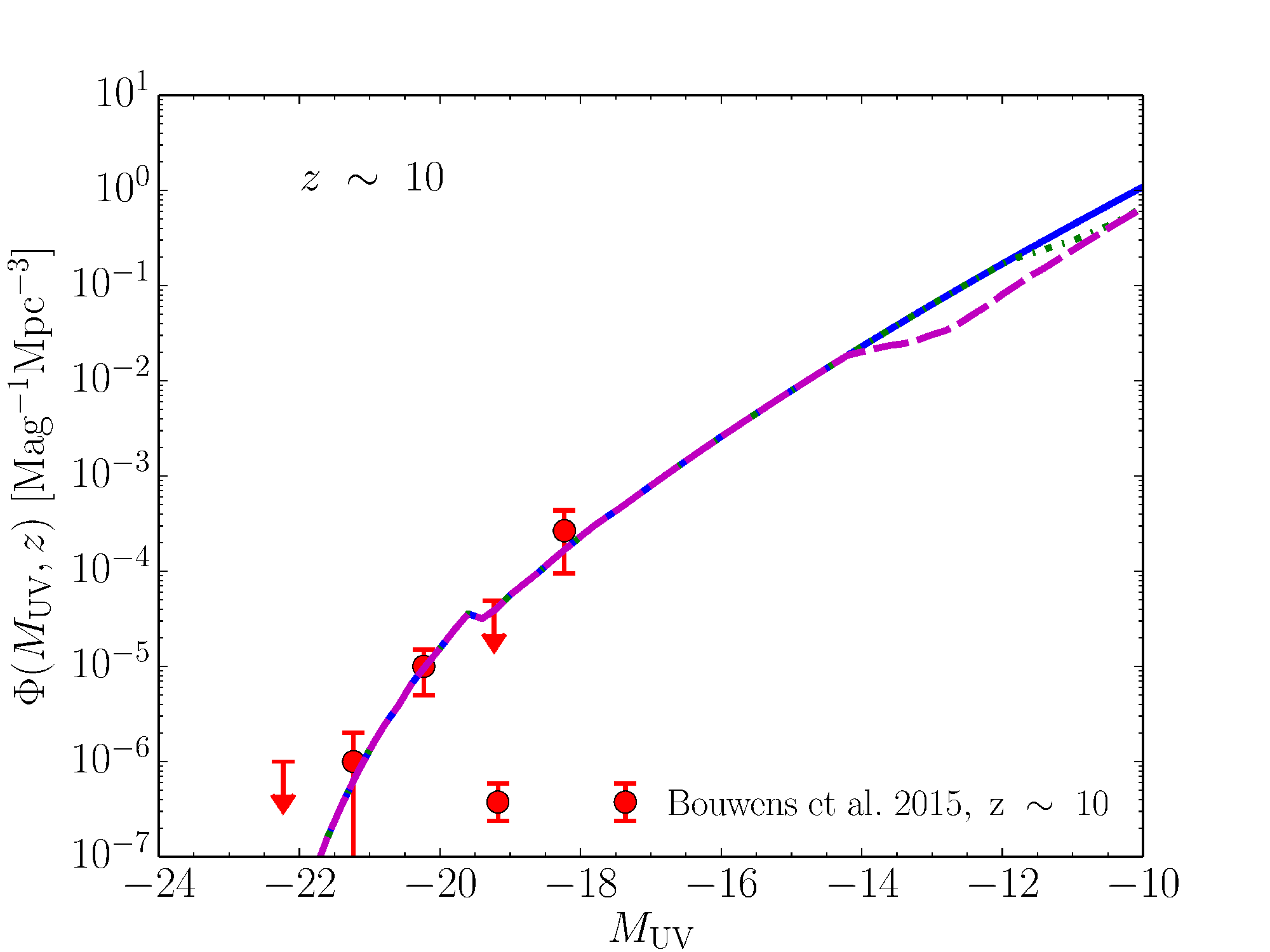}
\caption{The UV luminosity functions from our model with no feedback (blue), $v_c^{\rm cut}=50$ $km~s^{-1}$ (green) and $v_c^{\rm cut}=30$ $km~s^{-1}$ (magenta) at, from top to bottom, $z=5$, $z=8$ and $z=10$. Observed points are from \citet{2015ApJ...803...34B} (red), \citet{McLure2013} (green) and \citet{Atek2015} (magenta).}
\label{LF}
}
\end{figure}

For a given pair of parameters $f_{\rm esc}$ and $v_c^{\rm cut}$, we can now compute the probability $p_j(N^i_{\rm obs}|f_{\rm esc},v_c^{\rm cut})$ to observe a number of galaxies $N^i_{\rm obs}$ in the i-th magnitue bin following the j-th lensing model. We can then build the mean probability across all eight lensing models available in each field:
\begin{equation}
p(N^i_{\rm obs}|f_{\rm esc},v_c^{\rm cut}) = \frac{1}{8} \sum_j  p_j(N^i_{\rm obs}|f_{\rm esc},v_c^{\rm cut}),
\label{prob}
\end{equation}

As an example, we show $p$ as a function of $N^i$ in Fig.~\ref{probability} for the two fields A2744 and M0416 for the case when $f_{\rm esc} = 0.2$ and $v_c^{\rm cut} = 30$~km s$^{-1}$ compared for reference to the median of the number counts computed from the different lensing models. 

We compute the final likelhood assuming that the number counts in different magnitude bins are independent of each other. We also include an additional term to weight each model according to its consistency with the observed constraints on the CMB optical depth $\tau_{\rm obs}$:
\begin{equation}
L\propto{\rm exp}\left( \frac{-\chi_{\tau}^2}{2} \right)\times\prod_i  p(N^i_{\rm obs}|f_{\rm esc},v_c^{\rm cut}),
\label{LL}
\end{equation}
where 
\begin{equation}
\chi_{\tau}^2 = \frac{(\tau - \tau_{\rm obs})^2}{\sigma^2_\tau},
\end{equation}
$\tau_{\rm obs} = 0.0 66$ and $\sigma_\tau = 0 .016$ is the Planck measurement \citep{2015arXiv150201589P}, and $\tau$ is theoretical optical depth  which depends on the $f_{\rm esc}$ and $v_c^{\rm cut}$ \citep[e.g.][]{Shull2008,Yue2016}.

In Fig.~\ref{contour_z} we plot the contour map of the constraints on $f_{\rm esc}$ and $v_c^{\rm cut}$ from the combination of the two clusters A2744 and M0416. The number of ultra-faint FF galaxies yields a limit of $v_c^{\rm cut}\lesssim  50$ $km~s^{-1}$  (1$\sigma)$ on the cut-off circular velocity. In general terms, the number counts effectively constrain $v_c^{\rm cut}$, while no constraints can be put on $f_{\rm esc}$. Indeed, the dependence of the number counts on $f_{\rm esc}$ is mostly evident $\sim3-5$ magnitudes fainter than the LF turn-over magnitude at all redshifts, and/or at magnitudes close to the LF turn-over during reionization (i.e. $\gsim 7$), thus in a luminosity range which is not yet reached by current samples \citep[see][for details]{Yue2016}.
The constraint we obtain on $v_c^{\rm cut}$ can be translated into $\approx5.6\times10^9~M_\odot$ and  $\approx2.3\times10^9~M_\odot$ at $z=5$ and 10 respectively. In general, the smaller the halo mass is, the easier its star formation is quenched. Here what we get is the upper limit, above which one can safely say that halos can sustain continuous star formation. 
We verified that the inclusion in Eq.~\ref{LL} of the consistency criterion with the measured CMB optical depth has a minor effect on the above constraints. We show in Fig.~\ref{LF} the model UV luminosity functions at $z=5$, $8$, $10$ for reference $v_c^{\rm cut}$ values consistent with the limit we derived compared to the no-feedback case: the cut-off circular velocity corresponds to a UV cut-off which slightly depend on redshifts and roughly correspond to $M_{UV}\sim-15$ ($v_c^{\rm cut}=50$ $km~s^{-1}$) and $M_{UV}\sim-12$ ($v_c^{\rm cut}=30$ $km~s^{-1}$).
To improve these constraints and observe the intrinsic decline of galaxy abundance due to reionization feedback an improvement of either the observational data or of the lensing models is needed.
Our conclusions are robust against photometric redshift uncertainty: we found no appreciable change using different realizations of the high-$z$ sample obtained by randomly perturbing the photometric redshifts in the parent catalog according to the relevant uncertainty.

\begin{figure*}[!ht]
\centering{
\includegraphics[width=5.2cm]{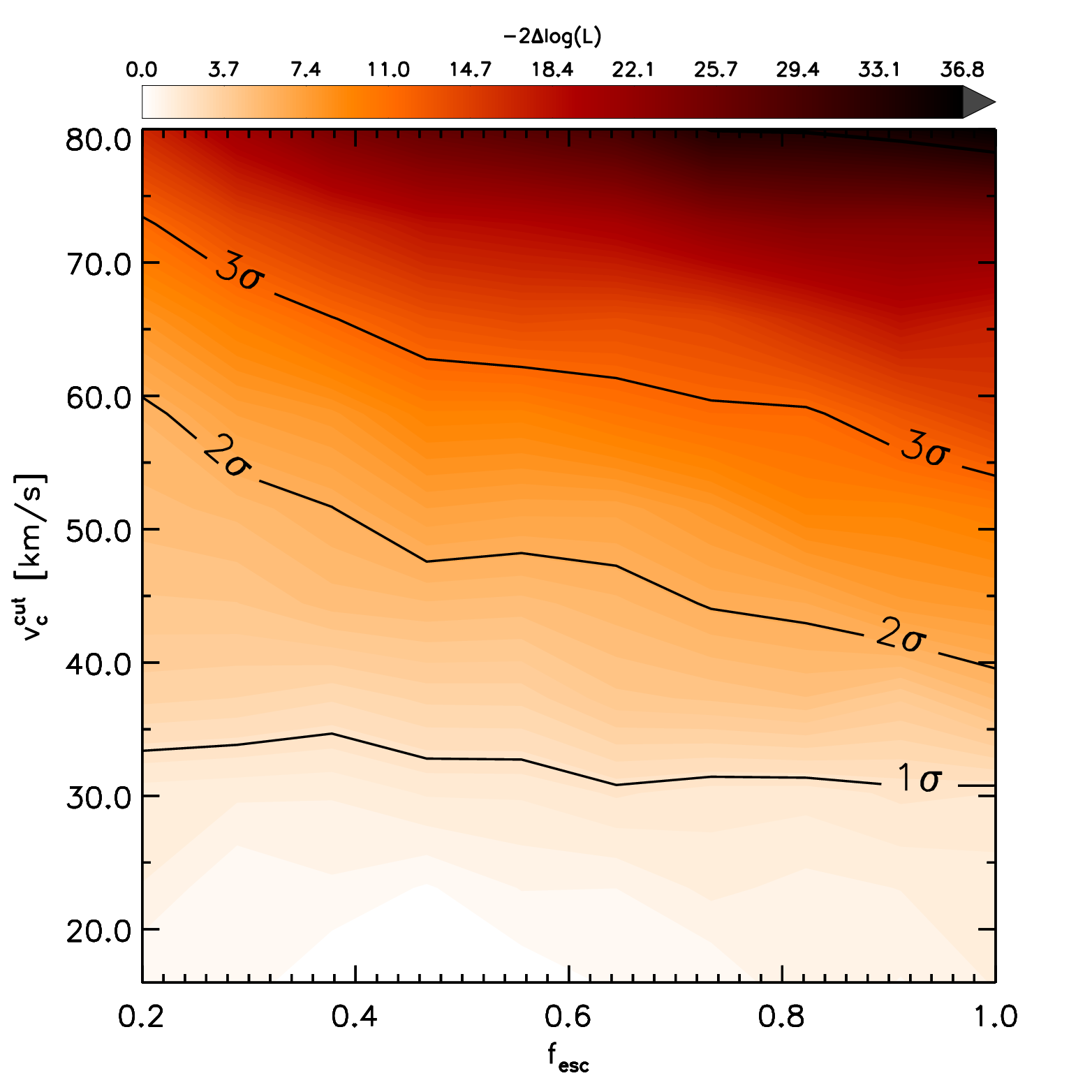}
\includegraphics[width=5.2cm]{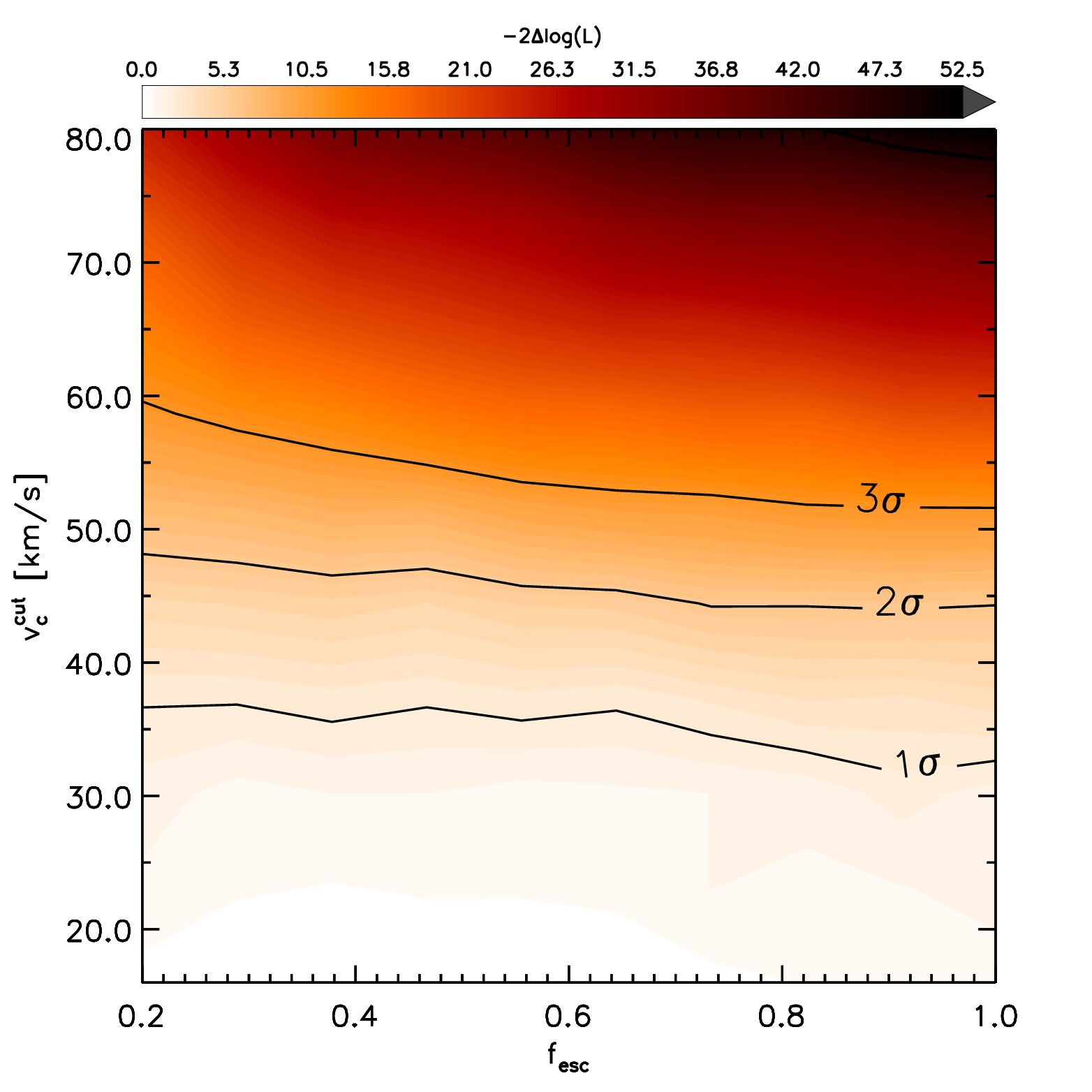}
\includegraphics[width=5.2cm]{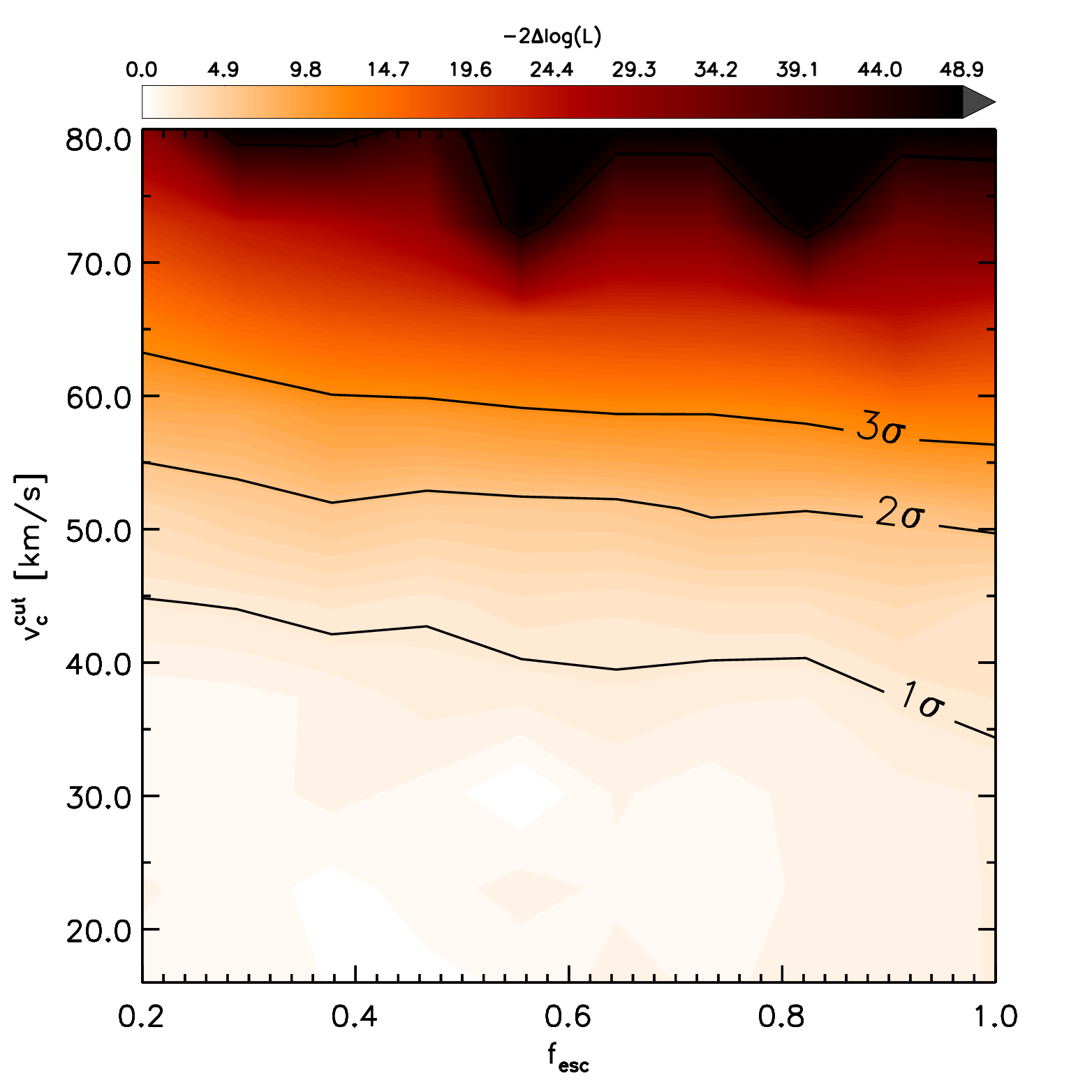}
 \caption{Parameter constraints obtained from three individual lensing models available for both clusters. From left to right: Brada\u{c}; Zitrin-nfw; Zitrin-ltm.
}
\label{contour_lens}
}
\end{figure*}

In Fig.~\ref{contour_lens} we show the contour maps obtained using three different lensing models that are available for both clusters.
On the one hand, individual models yield constraints that are in overall agreement with those outlined above. On the other hand, our ``global'' approach is more conservative since looser constraints than from individual maps are obtained when systematics are taken into account. This shows that improving lensing models and understanding their underlying discrepancies provides the best way to improve this kind of analysis.

\section{Summary and Conclusions}\label{sect_SUMMARY}

We have constrained our theoretical model (Sect.~\ref{sect_MODEL}) for the LF at high redshift using a sample of ultra-faint ($H_{160,int}>30$) $z>5$ galaxies in the first two Frontier Fields clusters A2744 and M0416. The objects have been selected on the basis of their photometric redshift computed from 10-bands photometry from the F435W to IRAC 4.5$\mu$m bands (Sect.~\ref{sect_FFDATA}). The comparison between theory and observations relies on the estimation of source detection completeness as a function of luminosity and size, and on taking into account systematics due to different lensing models. The free parameters of our model are the escape fraction of ionizing photons ($f_{\rm esc}$) and the cut-off circular velocity ($v_c^{\rm cut}$) below which star-formation is suppressed by photo-ionization feedback.
We find that galaxy number counts yield constraints on the reionization feedback strength while they are nearly unaffected by $f_{\rm esc}$. We found $v_c^{\rm cut} < 50$ $km~s^{-1}$, corresponding to a halo mass $\approx5.6\times10^9~M_\odot$ and  $\approx2.3\times10^9~M_\odot$ at $z=5$ and $10$ respectively and to $M_{UV}\approx-15$. Our analysis shows that photoionization feedback does not quench star formation activity in halos with circular velocity above 50 $km~s^{-1}$, while present data do not allow us to pinpoint the threshold below which feedback is effective. We find that the uncertainty in the lensing models, as embodied by systematic differences between different maps, is the factor that most limits our capability in putting stringent constraints on the effects of feedback on the high-redshift LF.

\acknowledgments
The research leading to these results has received funding from the European Union
Seventh Framework Programme (FP7/2007-2013) under grant agreement n. 312725. This work utilizes gravitational lensing models produced by PIs Brada\u{c}, Ebeling, Merten \& Zitrin, Sharon, and Williams funded as part of the HST Frontier Fields program conducted by STScI. STScI is operated by the Association of Universities for Research in Astronomy, Inc. under NASA contract NAS 5-26555. The lens models were obtained from the Mikulski Archive for Space Telescopes (MAST). A.M. and E.M.Q acknowledge funding from the STFC and a European Research Council 
Consolidator Grant (P.I. R. McLure). We thank R. Bouwens for useful discussions.


\begin{thebibliography}{43}
\expandafter\ifx\csname natexlab\endcsname\relax\def\natexlab#1{#1}\fi

\bibitem[{{Atek} {et~al.}(2015){Atek}, {Richard}, {Kneib}, {Jauzac},
  {Schaerer}, {Clement}, {Limousin}, {Jullo}, {Natarajan}, {Egami}, \&
  {Ebeling}}]{Atek2015}
{Atek}, H. {et~al.} 2015, \apj, 800, 18, 1409.0512

\bibitem[{{Barkana} {et~al.}(2001){Barkana}, {Haiman}, \&
  {Ostriker}}]{Barkana2001b}
{Barkana}, R., {Haiman}, Z., \& {Ostriker}, J.~P. 2001, \apj, 558, 482,
  astro-ph/0102304

\bibitem[{{Bouwens} {et~al.}(2015{\natexlab{a}}){Bouwens}, {Illingworth},
  {Oesch}, {Caruana}, {Holwerda}, {Smit}, \& {Wilkins}}]{Bouwens2015b}
{Bouwens}, R.~J., {Illingworth}, G.~D., {Oesch}, P.~A., {Caruana}, J.,
  {Holwerda}, B., {Smit}, R., \& {Wilkins}, S. 2015{\natexlab{a}}, \apj, 811,
  140, 1503.08228

\bibitem[{{Bouwens} {et~al.}(2015{\natexlab{b}}){Bouwens}, {Illingworth},
  {Oesch}, {Trenti}, {Labb{\'e}}, {Bradley}, {Carollo}, {van Dokkum},
  {Gonzalez}, {Holwerda}, {Franx}, {Spitler}, {Smit}, \&
  {Magee}}]{2015ApJ...803...34B}
{Bouwens}, R.~J. {et~al.} 2015{\natexlab{b}}, \apj, 803, 34, 1403.4295

\bibitem[{{Castellano} {et~al.}(2016{\natexlab{a}}){Castellano},
  {Amor{\'{\i}}n}, {Merlin}, {Fontana}, {McLure}, {M{\'a}rmol-Queralt{\'o}},
  {Mortlock}, {Parsa}, {Dunlop}, {Elbaz}, {Balestra}, {Boucaud}, {Bourne},
  {Boutsia}, {Brammer}, {Bruce}, {Buitrago}, {Capak}, {Cappelluti}, {Ciesla},
  {Comastri}, {Cullen}, {Derriere}, {Faber}, {Giallongo}, {Grazian}, {Grillo},
  {Mercurio}, {Michalowski}, {Nonino}, {Paris}, {Pentericci}, {Pilo}, {Rosati},
  {Santini}, {Schreiber}, {Shu}, \& {Wang}}]{Castellano2016b}
{Castellano}, M. {et~al.} 2016{\natexlab{a}}, ArXiv e-prints, 1603.02461

\bibitem[{{Castellano} {et~al.}(2016{\natexlab{b}}){Castellano}, {Dayal},
  {Pentericci}, {Fontana}, {Hutter}, {Brammer}, {Merlin}, {Grazian}, {Pilo},
  {Amorin}, {Cristiani}, {Dickinson}, {Ferrara}, {Gallerani}, {Giallongo},
  {Giavalisco}, {Guaita}, {Koekemoer}, {Maiolino}, {Paris}, {Santini},
  {Vallini}, {Vanzella}, \& {Wagg}}]{Castellano2016}
------. 2016{\natexlab{b}}, \apjl, 818, L3, 1601.03442

\bibitem[{{Coe} {et~al.}(2015){Coe}, {Bradley}, \& {Zitrin}}]{Coe2015}
{Coe}, D., {Bradley}, L., \& {Zitrin}, A. 2015, \apj, 800, 84, 1405.0011

\bibitem[{{Curtis-Lake} {et~al.}(2016){Curtis-Lake}, {McLure}, {Dunlop},
  {Rogers}, {Targett}, {Dekel}, {Ellis}, {Faber}, {Ferguson}, {Grogin},
  {Kocevski}, {Koekemoer}, {Lai}, {M{\'a}rmol-Queralt{\'o}}, \&
  {Robertson}}]{Curtis-Lake2016}
{Curtis-Lake}, E. {et~al.} 2016, \mnras, 457, 440, 1409.1832

\bibitem[{{Dayal} {et~al.}(2015){Dayal}, {Mesinger}, \& {Pacucci}}]{Dayal2015}
{Dayal}, P., {Mesinger}, A., \& {Pacucci}, F. 2015, \apj, 806, 67, 1408.1102

\bibitem[{{Dijkstra} {et~al.}(2004){Dijkstra}, {Haiman}, {Rees}, \&
  {Weinberg}}]{Dijkstra2004}
{Dijkstra}, M., {Haiman}, Z., {Rees}, M.~J., \& {Weinberg}, D.~H. 2004, \apj,
  601, 666, astro-ph/0308042

\bibitem[{{Finkelstein} {et~al.}(2015){Finkelstein}, {Ryan}, {Papovich},
  {Dickinson}, {Song}, {Somerville}, {Ferguson}, {Salmon}, {Giavalisco},
  {Koekemoer}, {Ashby}, {Behroozi}, {Castellano}, {Dunlop}, {Faber}, {Fazio},
  {Fontana}, {Grogin}, {Hathi}, {Jaacks}, {Kocevski}, {Livermore}, {McLure},
  {Merlin}, {Mobasher}, {Newman}, {Rafelski}, {Tilvi}, \&
  {Willner}}]{Finkelstein2015}
{Finkelstein}, S.~L. {et~al.} 2015, \apj, 810, 71, 1410.5439

\bibitem[{{Furlanetto} {et~al.}(2004){Furlanetto}, {Hernquist}, \&
  {Zaldarriaga}}]{2004MNRAS.354..695F}
{Furlanetto}, S.~R., {Hernquist}, L., \& {Zaldarriaga}, M. 2004, \mnras, 354,
  695, astro-ph/0406131

\bibitem[{{Galametz} {et~al.}(2013){Galametz}, {Grazian}, {Fontana},
  {Ferguson}, {Ashby}, {Barro}, {Castellano}, {Dahlen}, {Donley}, {Faber},
  {Grogin}, {Guo}, {Huang}, {Kocevski}, {Koekemoer}, {Lee}, {McGrath}, {Peth},
  {Willner}, {Almaini}, {Cooper}, {Cooray}, {Conselice}, {Dickinson}, {Dunlop},
  {Fazio}, {Foucaud}, {Gardner}, {Giavalisco}, {Hathi}, {Hartley}, {Koo},
  {Lai}, {de Mello}, {McLure}, {Lucas}, {Paris}, {Pentericci}, {Santini},
  {Simpson}, {Sommariva}, {Targett}, {Weiner}, {Wuyts}, \& {the CANDELS
  team}}]{Galametz2013}
{Galametz}, A. {et~al.} 2013, \apjs, 206, 10, 1305.1823

\bibitem[{{Giallongo} {et~al.}(2015){Giallongo}, {Grazian}, {Fiore}, {Fontana},
  {Pentericci}, {Vanzella}, {Dickinson}, {Kocevski}, {Castellano}, {Cristiani},
  {Ferguson}, {Finkelstein}, {Grogin}, {Hathi}, {Koekemoer}, {Newman}, \&
  {Salvato}}]{Giallongo2015}
{Giallongo}, E. {et~al.} 2015, \aap, 578, A83, 1502.02562

\bibitem[{{Guo} {et~al.}(2013){Guo}, {Ferguson}, {Giavalisco}, {Barro},
  {Willner}, {Ashby}, {Dahlen}, {Donley}, {Faber}, {Fontana}, {Galametz},
  {Grazian}, {Huang}, {Kocevski}, {Koekemoer}, {Koo}, {McGrath}, {Peth},
  {Salvato}, {Wuyts}, {Castellano}, {Cooray}, {Dickinson}, {Dunlop}, {Fazio},
  {Gardner}, {Gawiser}, {Grogin}, {Hathi}, {Hsu}, {Lee}, {Lucas}, {Mobasher},
  {Nandra}, {Newman}, \& {van der Wel}}]{Guo2013}
{Guo}, Y. {et~al.} 2013, \apjs, 207, 24

\bibitem[{{Hasegawa} \& {Semelin}(2013)}]{2013MNRAS.428..154H}
{Hasegawa}, K., \& {Semelin}, B. 2013, \mnras, 428, 154, 1209.4143

\bibitem[{{Huang} {et~al.}(2013){Huang}, {Ferguson}, {Ravindranath}, \&
  {Su}}]{2013ApJ...765...68H}
{Huang}, K.-H., {Ferguson}, H.~C., {Ravindranath}, S., \& {Su}, J. 2013, \apj,
  765, 68, 1301.4443

\bibitem[{{Ishigaki} {et~al.}(2015){Ishigaki}, {Kawamata}, {Ouchi}, {Oguri},
  {Shimasaku}, \& {Ono}}]{Ishigaki2015}
{Ishigaki}, M., {Kawamata}, R., {Ouchi}, M., {Oguri}, M., {Shimasaku}, K., \&
  {Ono}, Y. 2015, \apj, 799, 12, 1408.6903

\bibitem[{{Khaire} {et~al.}(2015){Khaire}, {Srianand}, {Choudhury}, \&
  {Gaikwad}}]{Khaire2015}
{Khaire}, V., {Srianand}, R., {Choudhury}, T.~R., \& {Gaikwad}, P. 2015, ArXiv
  e-prints, 1510.04700

\bibitem[{{Laporte} {et~al.}(2014){Laporte}, {Streblyanska}, {Clement},
  {P{\'e}rez-Fournon}, {Schaerer}, {Atek}, {Boone}, {Kneib}, {Egami},
  {Mart{\'{\i}}nez-Navajas}, {Marques-Chaves}, {Pell{\'o}}, \&
  {Richard}}]{Laporte2014}
{Laporte}, N. {et~al.} 2014, \aap, 562, L8, 1401.8263

\bibitem[{{Ma} {et~al.}(2016){Ma}, {Hopkins}, {Kasen}, {Quataert},
  {Faucher-Giguere}, {Keres}, \& {Murray}}]{Ma2016}
{Ma}, X., {Hopkins}, P.~F., {Kasen}, D., {Quataert}, E., {Faucher-Giguere},
  C.-A., {Keres}, D., \& {Murray}, N. 2016, ArXiv e-prints, 1601.07559

\bibitem[{{Mason} {et~al.}(2015){Mason}, {Trenti}, \&
  {Treu}}]{2015arXiv150801204M}
{Mason}, C., {Trenti}, M., \& {Treu}, T. 2015, ArXiv e-prints, 1508.01204

\bibitem[{{McLeod} {et~al.}(2015){McLeod}, {McLure}, {Dunlop}, {Robertson},
  {Ellis}, \& {Targett}}]{McLeod2015}
{McLeod}, D.~J., {McLure}, R.~J., {Dunlop}, J.~S., {Robertson}, B.~E., {Ellis},
  R.~S., \& {Targett}, T.~A. 2015, \mnras, 450, 3032, 1412.1472

\bibitem[{{McLure} {et~al.}(2013){McLure}, {Dunlop}, {Bowler}, {Curtis-Lake},
  {Schenker}, {Ellis}, {Robertson}, {Koekemoer}, {Rogers}, {Ono}, {Ouchi},
  {Charlot}, {Wild}, {Stark}, {Furlanetto}, {Cirasuolo}, \&
  {Targett}}]{McLure2013}
{McLure}, R.~J. {et~al.} 2013, \mnras, 432, 2696, 1212.5222

\bibitem[{{Menci} {et~al.}(2016){Menci}, {Sanchez}, {Castellano}, \&
  {Grazian}}]{Menci2016}
{Menci}, N., {Sanchez}, N.~G., {Castellano}, M., \& {Grazian}, A. 2016, \apj,
  818, 90, 1601.01820

\bibitem[{{Merlin} {et~al.}(2016){Merlin}, {Amor{\`i}n}, {Castellano},
  {Fontana}, {Buitrago}, {Dunlop}, {Elbaz}, {Boucaud}, {Bourne}, {Boutsia},
  {Brammer}, {Bruce}, {Capak}, {Cappelluti}, {Ciesla}, {Comastri}, {Cullen},
  {Derriere}, {Faber}, {Ferguson}, {Giallongo}, {Grazian}, {Lotz},
  {Michalowski}, {Paris}, {Pentericci}, {Pilo}, {Santini}, {Schreiber}, {Shu},
  \& {Wang}}]{Merlin2016}
{Merlin}, E. {et~al.} 2016, ArXiv e-prints, 1603.02460

\bibitem[{{Merlin} {et~al.}(2015){Merlin}, {Fontana}, {Ferguson}, {Dunlop},
  {Elbaz}, {Bourne}, {Bruce}, {Buitrago}, {Castellano}, {Schreiber}, {Grazian},
  {McLure}, {Okumura}, {Shu}, {Wang}, {Amor{\'{\i}}n}, {Boutsia}, {Cappelluti},
  {Comastri}, {Derriere}, {Faber}, \& {Santini}}]{Merlin2015}
------. 2015, \aap, 582, A15, 1505.02516

\bibitem[{{Mesinger} \& {Dijkstra}(2008)}]{2008MNRAS.390.1071M}
{Mesinger}, A., \& {Dijkstra}, M. 2008, \mnras, 390, 1071, 0806.3090

\bibitem[{{Oesch} {et~al.}(2014){Oesch}, {Bouwens}, {Illingworth}, {Franx},
  {Ammons}, {van Dokkum}, {Trenti}, \& {Labbe}}]{Oesch2014}
{Oesch}, P.~A., {Bouwens}, R.~J., {Illingworth}, G.~D., {Franx}, M., {Ammons},
  S.~M., {van Dokkum}, P.~G., {Trenti}, M., \& {Labbe}, I. 2014, ArXiv
  e-prints, 1409.1228

\bibitem[{{Planck Collaboration} {et~al.}(2015){Planck Collaboration}, {Ade},
  {Aghanim}, {Arnaud}, {Ashdown}, {Aumont}, {Baccigalupi}, {Banday},
  {Barreiro}, {Bartlett}, \& et~al.}]{2015arXiv150201589P}
{Planck Collaboration} {et~al.} 2015, ArXiv e-prints, 1502.01589

\bibitem[{{Robertson} {et~al.}(2015){Robertson}, {Ellis}, {Furlanetto}, \&
  {Dunlop}}]{Robertson2015}
{Robertson}, B.~E., {Ellis}, R.~S., {Furlanetto}, S.~R., \& {Dunlop}, J.~S.
  2015, \apjl, 802, L19, 1502.02024

\bibitem[{{Sharma} {et~al.}(2016){Sharma}, {Theuns}, {Frenk}, {Bower}, {Crain},
  {Schaller}, \& {Schaye}}]{Sharma2016}
{Sharma}, M., {Theuns}, T., {Frenk}, C., {Bower}, R., {Crain}, R., {Schaller},
  M., \& {Schaye}, J. 2016, \mnras, 1512.04537

\bibitem[{{Shull} \& {Venkatesan}(2008)}]{Shull2008}
{Shull}, J.~M., \& {Venkatesan}, A. 2008, \apj, 685, 1, 0806.0392

\bibitem[{{Sobacchi} \& {Mesinger}(2013{\natexlab{a}})}]{2013MNRAS.432.3340S}
{Sobacchi}, E., \& {Mesinger}, A. 2013{\natexlab{a}}, \mnras, 432, 3340,
  1301.6781

\bibitem[{{Sobacchi} \& {Mesinger}(2013{\natexlab{b}})}]{2013MNRAS.432L..51S}
------. 2013{\natexlab{b}}, \mnras, 432, 51, 1301.6776

\bibitem[{{Stanway} {et~al.}(2016){Stanway}, {Eldridge}, \&
  {Becker}}]{Stanway2016}
{Stanway}, E.~R., {Eldridge}, J.~J., \& {Becker}, G.~D. 2016, \mnras, 456, 485,
  1511.03268

\bibitem[{{Tacchella} {et~al.}(2013){Tacchella}, {Trenti}, \&
  {Carollo}}]{2013ApJ...768L..37T}
{Tacchella}, S., {Trenti}, M., \& {Carollo}, C.~M. 2013, \apjl, 768, L37,
  1211.2825

\bibitem[{{Trenti} {et~al.}(2010){Trenti}, {Stiavelli}, {Bouwens}, {Oesch},
  {Shull}, {Illingworth}, {Bradley}, \& {Carollo}}]{2010ApJ...714L.202T}
{Trenti}, M., {Stiavelli}, M., {Bouwens}, R.~J., {Oesch}, P., {Shull}, J.~M.,
  {Illingworth}, G.~D., {Bradley}, L.~D., \& {Carollo}, C.~M. 2010, \apjl, 714,
  L202, 1004.0384

\bibitem[{{Yoshiura} {et~al.}(2016){Yoshiura}, {Hasegawa}, {Ichiki}, {Tashiro},
  {Shimabukuro}, \& {Takahashi}}]{Yoshiura2016}
{Yoshiura}, S., {Hasegawa}, K., {Ichiki}, K., {Tashiro}, H., {Shimabukuro}, H.,
  \& {Takahashi}, K. 2016, ArXiv e-prints, 1602.04407

\bibitem[{{Yue} {et~al.}(2014){Yue}, {Ferrara}, {Vanzella}, \&
  {Salvaterra}}]{2014MNRAS.443L..20Y}
{Yue}, B., {Ferrara}, A., {Vanzella}, E., \& {Salvaterra}, R. 2014, \mnras,
  443, L20, 1405.3440

\bibitem[{{Yue} {et~al.}(2016){Yue}, {Ferrara}, \& {Xu}}]{Yue2016}
{Yue}, B., {Ferrara}, A., \& {Xu}, Y. 2016, ArXiv e-prints, 1604.01314

\bibitem[{{Zheng} {et~al.}(2014){Zheng}, {Shu}, {Moustakas}, {Zitrin}, {Ford},
  {Huang}, {Broadhurst}, {Molino}, {Diego}, {Infante}, {Bauer}, {Kelson}, \&
  {Smit}}]{Zheng2014}
{Zheng}, W. {et~al.} 2014, \apj, 795, 93, 1402.6743

\bibitem[{{Zitrin} {et~al.}(2014){Zitrin}, {Zheng}, {Broadhurst}, {Moustakas},
  {Lam}, {Shu}, {Huang}, {Diego}, {Ford}, {Lim}, {Bauer}, {Infante}, {Kelson},
  \& {Molino}}]{Zitrin2014}
{Zitrin}, A. {et~al.} 2014, \apjl, 793, L12, 1407.3769

\end{thebibliography}
\end{document}